\documentclass{article}

\usepackage{microtype}
\usepackage{graphicx}
\usepackage{subfigure}
\usepackage{booktabs} 
\usepackage{amsmath}
\usepackage{amsfonts}

\usepackage{hyperref}

\makeatletter
\newcommand\footnoteref[1]{\protected@xdef\@thefnmark{\ref{#1}}\@footnotemark}
\makeatother

\usepackage[accepted]{icml2018}

\icmltitlerunning{BachProp: Learning to Compose Music in Multiple Styles}

\begin{document}

\twocolumn[
\icmltitle{BachProp: Learning to Compose Music in Multiple Styles}



\icmlsetsymbol{equal}{*}

\begin{icmlauthorlist}
\icmlauthor{Florian Colombo}{lcn1,lcn2}
\icmlauthor{Wulfram Gerstner}{lcn1,lcn2}
\end{icmlauthorlist}

\icmlaffiliation{lcn1}{Brain-Mind Institute, School of Life Sciences, \'Ecole Polytechnique F\'ed\'erale de Lausanne, Lausanne, Switzerland}
\icmlaffiliation{lcn2}{School of Computer and Communication
Sciences, \'Ecole Polytechnique F\'ed\'erale de Lausannee, Lausanne, Switzerland}

\icmlcorrespondingauthor{Florian Colombo}{florian.colombo@epfl.ch}

\icmlkeywords{Music, LSTM}

\vskip 0.3in
]



\printAffiliationsAndNotice{}  

\begin{abstract}
Hand in hand with deep learning advancements, algorithms of music composition increase in performance. However, most of the successful models are designed for specific musical structures. Here, we present BachProp, an algorithmic composer that can generate music scores in any style given sufficient training data. To adapt BachProp to a broad range of musical styles, we propose a novel normalized representation of music and train a deep network to predict the note transition probabilities of a given music corpus. In this paper, new music scores sampled by BachProp are compared with the original corpora via crowdsourcing. This evaluation indicates that the music scores generated by BachProp are not less preferred than the original music corpus the algorithm was provided with. 

\end{abstract}

\section{Introduction}
\label{introduction}



In search of the computational creativity frontier \cite{colton2012computational}, machine learning algorithms are more and more present in creative domains such as painting \cite{mordvintsev2015inceptionism,gatys2016image} and music \cite{sturm2016music, colombo2017deep, hadjeres2016deepbach}. Already in 1847, Ada Lovelace predicted the potential of analytical engines for algorithmic music composition \cite{lovelace1843notes}. Current methods include rule based approaches, genetic algorithms, Markov models or more recently artificial neural networks \cite{fernandez2013ai}. 

One of the first artificial neural networks applied to music composition was a recurrent neural network  trained to generate monophonic melodies \cite{todd1989connectionist}. The long short-term memory (LSTM) networks introduced by \cite{hochreiter1997long} was first applied to music composition, so as to generate Blues monophonic melodies constrained on chord progressions \cite{eck2002finding}. Since then, music composition algorithms employing LSTM units, have been used to generate monophonic melodies of folk music \cite{sturm2016music, colombo2017deep} or chorales harmonized in the style of Bach  \cite{liang2016bachbot, hadjeres2016deepbach}. However, most of these algorithms make strong assumptions about the structure of the music they modeled. For example, \cite{sturm2016music, colombo2017deep} are designed for monophonic melodies only. \cite{eck2002finding} can only generate monophonic melodies on top of simple chords with a fixed rhythm. The works of \cite{liang2016bachbot} and \cite{hadjeres2016deepbach} focused on the task of harmonization in the style of Bach and therefore designed algorithms that exhibit and inductive bias toward the structure of the chorales. In addition, they require preprocessing of the data set that cannot easily be adapted to other music data in an automatized way. 

Here, we present a neural composer algorithm named \emph{BachProp} designed to generate new music scores in any style. To this end, we do not assume any specific musical structure of the data except that it is composed of sequences of notes (belonging to the 12-tone system). The Musical Instrument Digital Interface (MIDI) is a symbolic notation of music that was designed to record digital synthesizers. Because MIDI files have potentially infinitely many different ways of representing the same music score, especially rhythm, we developed a novel normalized representation of music that can be applied to most MIDI sequences with minimal distortion. 

BachProp is a generative model of music scores that learns the musical structure from MIDI sequences. With the normalized MIDI representation together with the architecture, training and generation methods presented in this paper, BachProp is able to create new music scores with composition structures extracted from any relatively large data set. Each piece can last for an arbitrary time during which it exhibits well-defined and consistent features. For example, when trained on string quartets from Mozart and Haydn, BachProp can generate music scores mimicking the composers' style in any key and each with a unique mood. Thanks to the normalized representation of MIDI sequences, BachProp can be trained on data sets previously unused by deep learning algorithms such as the recordings of Bach keyboard works by John Sankey on a digital keyboard. 


In Section \ref{sec:midi}, we present and motivate the normalized MIDI representation. Section \ref{model} introduces BachProp's neural network along with the training and generation procedures. In section \ref{sec:results}, we analyze the predictive and generation performance of BachProp on five data sets with different levels of heterogeneity of musical structure. Finally, we discuss and present a challenge designed to compare the music generated from automated and human composers in section \ref{evaluation}.




\section{Normalized MIDI representation}
\label{sec:midi}


\begin{figure}
\includegraphics[width=\columnwidth]{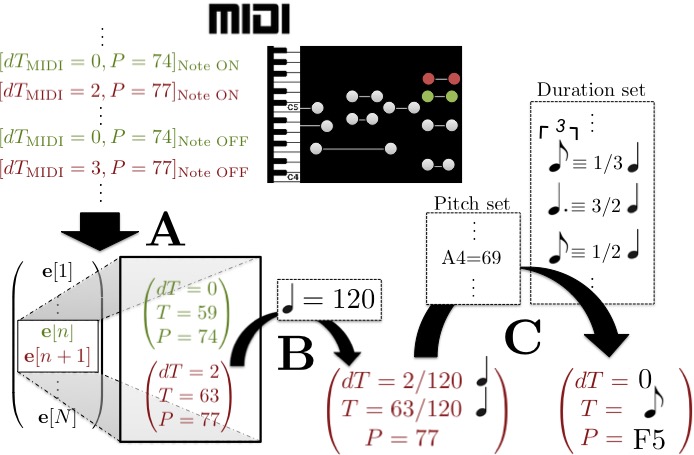}
\caption{\textbf{Normalized MIDI representation.} An illustration of the steps involved in the proposed encoding of MIDI sequences. Using two sets that defines the possible durations and pitches, a MIDI sequence (top left) is translated in a sequence of notes with three features: the timing $dT$, the duration $T$ and the pitch $P$.}
\label{fig:midi}
\end{figure}

MIDI is a protocol which was designed for digital synthesizers. It is, just as standard music notation, a symbolic representation of music. However, because it was designed to record and play back rather than to write scores, standard music composition softwares such as MuseScore have difficulties to correctly and consistently translate the content of MIDI files into the standard music notation. The difficulty lies not in determining the pitch $P$ of each note (an integer between 0 and 126 where A4=69) but in inferring the correct sequence of durations in terms of the commonly used notation for rhythm (e.g. quarter note or eighth note triplet). We aim to solve this problem by normalizing MIDI sequences to a shared and low dimensional representation. In Section \ref{model}, we then use the normalized music score as input to a deep LSTM network. 


There has been several approaches to represent polyphonic MIDI sequences in neural networks. The encoding methods of \cite{boulanger2012modeling,liang2016bachbot,hadjeres2016deepbach} all discretize time into frames of fixed duration (e.g. 16th note). While this representation works on rare data sets as Bach chorales where all note duration are a multiple of this base duration, it cannot be applied to most MIDI sequences without distorting rhythms. The method described below and illustrated in Figure \ref{fig:midi} allows to represent every MIDI file (including polyphony and multitrack) in term of a sequence of notes in a controlled rhythmic environment. Similar approaches were taken for monophonic melodies from MIDI \cite{colombo2017deep} or higher level music notation languages such as the ABC notation \cite{sturm2016music}.

%

A MIDI file contains a header (meta parameters) and possibly multiple tracks defined each by a sequence of MIDI messages. For BachProp, we are only interested in the MIDI messages defining when a note is being pressed down (MIDI note ON events) or released (MIDI note OFF events). In MIDI, the precise timing of the $m$th event is encoded relatively to the previous event (at position $m-1$). As a consequence, every MIDI message shares the common attribute $dT[m]$, which represents the number of ticks that separates each message from the previous. MIDI ticks are natural numbers that directly relate to time. For example, in a MIDI sequence with 120 beats per minute (BPM) and 192 pulses per quarter notes (PPQ), the duration of one tick is 
\begin{equation}
\frac{60000 \text{ [ms/min]}}{120 \text{ [beat/min]} \times 192 \text{ [tick/beat]}} = 2.604 \text{ [ms/tick]}.
\end{equation}
The PPQ and BPM are defined in the header of every MIDI file. Note that MIDI messages modulating the BPM can be sent at any time in a MIDI sequence. 

\subsection{Extracting notes from MIDI}

Towards a normalized representation of music, the first step involves parsing the list of messages to extract for the same note, the ON and OFF events. Then, we merge the two consecutive events involving the same note pitch $P$ to translate sequence of MIDI messages into a sequence of notes (Figure \ref{fig:midi}\textbf{A}). In addition, we transform the MIDI tick durations to quarter note durations (Figure \ref{fig:midi}\textbf{B}). In the resulting sequence of notes, each note is represented by 
\begin{equation}
\mathbf{note}[n] = (dT[n], T[n], P[n]).
\label{eq:event}
\end{equation}
$T[n]$ and $P[n]$ represents duration and pitch of a note, while $dT[n]$ is the timing of the $n$th note with respect to the previous one. A value $dT[n]=0$ indicates synchronous chords. 

\subsection{Rhythm normalization} 

MIDI being designed to record live performances on digital keyboards, many MIDI files exhibit expressive and rhythmic freedom that leads to small variations around the actual duration of the note in the standard music notation. In addition, the zoo of all MIDI sequences on the internet was created by many different softwares, each with their own way of representing standard music durations. Therefore, we perform a final normalization step (Figure \ref{fig:midi}\textbf{C}) towards a low-dimensional encoding of MIDI sequences. We map all timings and durations to a set of 21 possible note lengths (duration set) expressed as fractions or multiples of quarter notes, similar to durations in standard music notation softwares. Mapping to the closest value in the set (Euclidian distance) removes the temporal jittering around the original note duration. The result is $dT[n]$ and $T[n]$ being constrained to a discrete set of values.

\begin{figure}
\begin{tabular}{l}
\bf{A}\\
\includegraphics[width=\columnwidth]{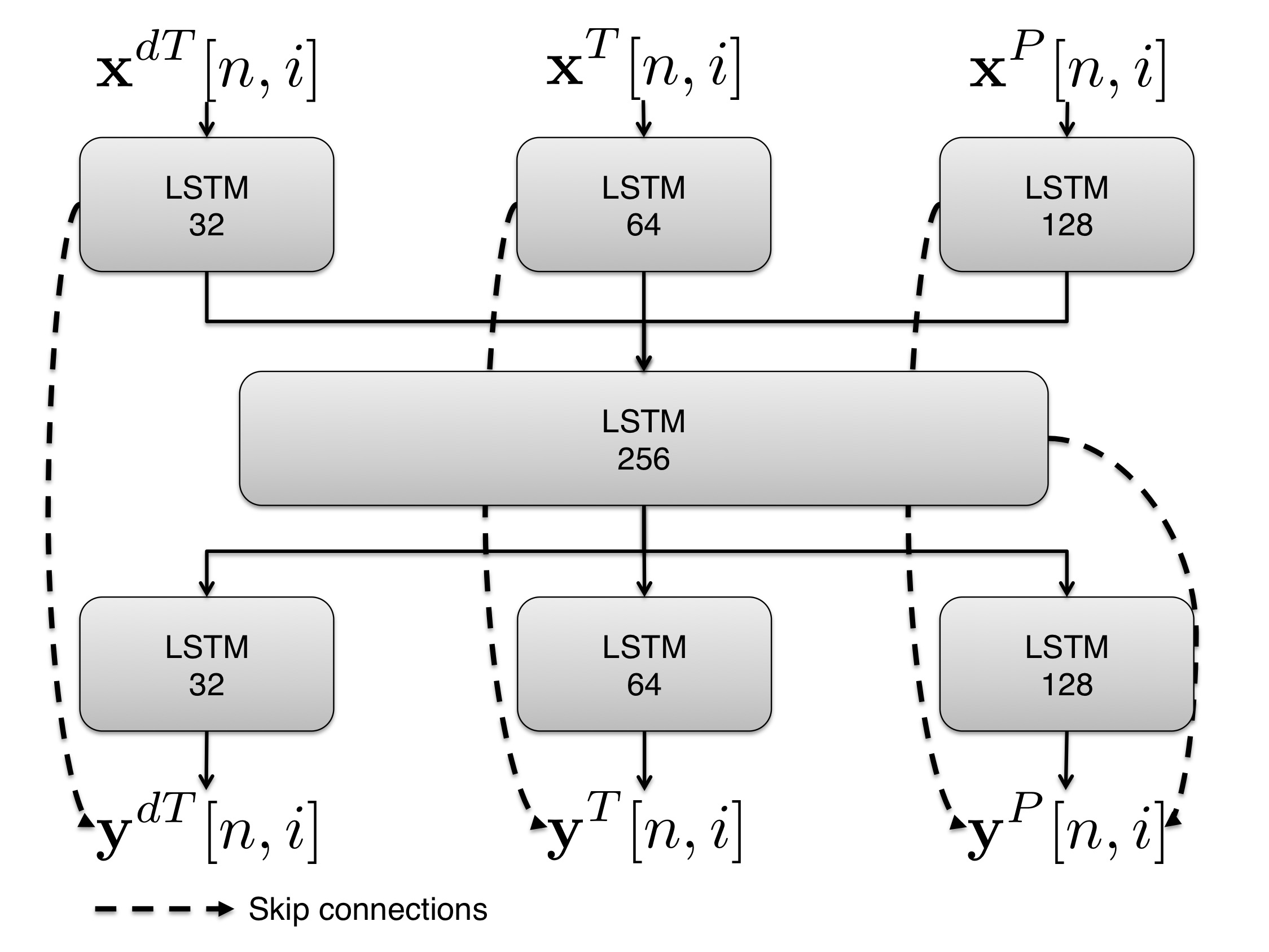}\\
\bf{B}\\
\includegraphics[width=\columnwidth]{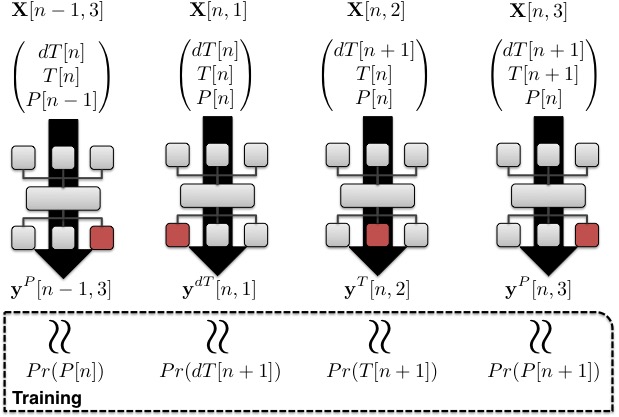}
\end{tabular}
\caption{\textbf{Neural network architecture.} \textbf{A} Schematic of BachProp network architecture. Blocks of LSTM units are connected in a feedforward structure (solid and dashed arrows). Each block contains 32, 64, 128, or 256 fully connected LSTM units, as indicated. Feedforward connections have a dropout with a probability of 0.3. The output layer depicted by $\mathbf{y}^F$ is a softmax operator over output units used to represent the probability of possible realisation of the next note feature $F \in [dT,T,P]$. \textbf{B} Illustration of input and output vectors. $\mathbf{X}[n,i]$ is a list of the three input vectors $\mathbf{x}^{dT}$, $\mathbf{x}^T$ and $\mathbf{x}^P$ in this order at time step $(n,i)$. These vectors use one-hot encoding of the corresponding note feature. The network is trained to approximate the transition probabilities $Pr(\mathbf{X}[n+1]|\mathbf{X}[n],\mathbf{H}[n])$, where $\mathbf{H}$ is the hidden state of recurrent units. The arrows depict the non linear transformation of the input through the neural network of Figure \ref{fig:architecture}\textbf{A}. The result of this operation are the output vectors $\mathbf{y}^F$. For each note $n$, the procedure cycles through 3 substeps to predict timing $dT$, duration $T$, and pitch $P$ (red output boxes).}
\label{fig:architecture}
\end{figure}

\section{BachProp}
\label{model}

BachProp is a generative model of MIDI sequences. It combines the normalized MIDI representation, a recurrent neural network architecture with specific inputs and outputs at each time step, a parameter optimization procedure and finally a method to generate new music scores from the trained model. 

\subsection{Network architecture}

We used a deep LSTM \cite{hochreiter1997long} network with three consecutive layers as schematized in Figure \ref{fig:architecture}\textbf{A}. The network's task is to infer the probability distribution over the next possible notes from the representation of the current note and the network's internal state (the network representation of the history of notes). To facilitate gradient propagation, we added skip connections. Note that we employ more units for melody (represented by the pitch  $P$) than for rhythm (represented by the timing $dT$ and duration $T$). 

In most western music, there exists a dependence between a note duration and its rhythm. For this reason, we unrolled in time the predictions of timing, duration, and pitch. In particular, the prediction of the upcoming pitch is conditioned not only on the current note, but also on the upcoming timing and duration (Figure \ref{fig:architecture}\textbf{B}). The underlying probability distributions modeled by the artificial neural network of BachProp are detailed in the next section.

\subsection{Probability of note sequences}

The artificial neural network from which BachProp draws new samples is trained to approximate the probability distributions of music scores by minimizing the negative log-likelihood (NLL) 
\begin{align}
\mathcal{L} &= \sum_{s=1}^S \log\big(Pr(\text{song}_s)\big) \nonumber\\
&= \sum_{s=1}^S \log\big(Pr(\mathbf{note}_s[1:N_s])),
\end{align}

The probability of song $s$ with $N_s$ notes is defined as the joint probability of all notes. This probability can be rewritten with conditional probabilities
\begin{align}
 Pr(&\mathbf{note}_s[1:N_s]) = \nonumber\\
 &Pr(\mathbf{note}_s[1])\prod_{n=1}^{N_s-1}Pr(\mathbf{note}_s[n+1]|\mathbf{note}_s[1:n]).
 \label{eq:prS}
\end{align}
In our recurrent neural network, the history of events is represented by the hidden state $\mathbf{H}[n]$ of recurrent units. Consequently, the output $\mathbf{y}[n]$ of the network after $n$ time steps is a non linear transform $\Phi$ of the input - the representation of the note event $n$ - and the hidden state. The output is used to predict the next note event.
\begin{align}
 \mathbf{y}[n] &= \Phi(\mathbf{note}[n],\mathbf{H}[n])\nonumber\\
  &\approx  Pr(\mathbf{note}[n+1]|\mathbf{note}[1:n]).
 \label{eq:y}
\end{align}

Importantly, the probability of an event corresponds to the joint probability of all three features of this event: timing $dT$, duration $T$, and pitch $P$. If we consider timing and duration sets with 21 entries and a pitch set containing all 88 piano keys, this joint distribution has $21\times21\times88=38808$ dimensions. To reduce this space and account for the relationship between the note timing, duration and pitch features, we split the joint probability of the three features of an event into a product of conditional probabilities:
\begin{align}
Pr&(dT[n], T[n], P[n]||\mathbf{e}[1:n]) = \nonumber\\ 
&Pr(dT[n+1]|\mathbf{e}[1:n]) \times\nonumber\\ 
&Pr(T[n+1]|\mathbf{e}[1:n],dT[n+1])\times\nonumber\\ 
&Pr(P[n+1]|\mathbf{e}[1:n],T[n+1],dT[n+1]).
\label{eq:prE}
\end{align}
In words, we predict the pitch of $\mathbf{note}[n+1]$ only after choosing its timing and duration. In our network architecture, we implement this idea by splitting each time step $n$ in three substeps (see Figure \ref{fig:architecture}\textbf{B}). Effectively, the timing distribution $Pr(dT[n+1]|\mathbf{note}[1:n])$ is read out after the first substep ($i$=1), where the input is the representation of the previous note. Then, the probability distribution $Pr(T[n+1]|\mathbf{note}[1:n],dT[n+1])$ is read out at the second substep ($i$=2), where the upcoming timing is provided as timig input $\mathbf{x}^{dT}[n,2]$. Finally, the pitch probability distribution $Pr(P[n+1]|\mathbf{note}[1:n],T[n+1],dT[n+1])$ is read out at the final substep ($i$=3), where the upcoming timing and duration are provided as inputs. While during training, the timing and duration of the upcoming notes are known, for generation we sample from the probability distributions $Pr(dT[n+1])$ and $Pr(T[n+1])$ to proceed through the three substeps.


\subsection{Training}
\label{sec:training}

The artificial network of BachProp is trained to approximate the probability distributions shown in Figure \ref{fig:architecture}\textbf{B}. To achieve this, we perform 2000 consecutive rounds of training (epochs). In each epoch, all songs in the training corpus (90\% of the original data set) are presented to the network. Backpropagation through time is applied to compute the error signal. The parameter updates are performed on batches of 32 songs with the Adam optimizer \cite{kingma2014adam} and the NLL loss function. To prevent exploding gradients, we clip the norm of the gradient and evaluate the backpropagation signal on sequences of 64 consecutive notes. The network hidden state is reset at the beginning of songs only. At the end of each epoch, we evaluate the predictive performance of the network on the remaining 10\% of the original music corpus. The parameters that maximized the accuracy are saved as optimal parameters.  

Using our design (see Figure \ref{fig:architecture}\textbf{B}), the conditional distributions are read out and used at different substep $i$ corresponding to a feature: $i=1$ for $Pr(dT[n+1])$, $i=2$ for $Pr(T[n+1])$, and $i=3$ for $Pr(P[n+1])$. However, the network is trained to approximate these distributions already two time steps before. Effectively learning a representation that hints toward the desired output. Taking pitch prediction as example, the network is trained to predict the upcoming pitch $Pr(P[n+1])$ based first on the last note event only, then additionally knowing the relative timing $dT[n+1]$ and finally based on the duration $T[n+1]$ as well. Therefore, the network tracks not only the temporal structure of notes but also the relationship between the features constituting notes. 

\subsubsection{Data augmentation}
In order for BachProp to learn tonality and transposition invariance of music, we randomly transpose each song at the beginning of every training epoch within the available bounds of the pitch set. In other words, for each song we compute the possible shifts of semitones (basic musical unit separating each pitch) and sample one that is then applied as an offset to all pitches in the song. Because a single MIDI sequence will be transposed with up to 50 offsets, this augmentation method allows BachProp to learn the temporal structure of music on more examples.

\subsection{Music generation} 

Once trained, the output probabilities of the neural network are sampled to select each upcoming note being generated by BachProp. We start by randomly selecting the first note only and feed it to the network as $\mathbf{X}[1,1]$. We then iteratively sample $dT[n+1]$, $T[n+1]$ and $P[n+1]$ from the network output probabilities at the corresponding time step $n$ and substep $i$. This generated note is given as input to the network for the next time step. The resulting sequence of notes in our representation can easily be translated back to a MIDI sequence by reversing the method schematized in Figure \ref{fig:midi}, except that we do not add jitter.


To sample the probability distributions, we constrain the possible realizations to the $M$ most likely ones. Nonetheless, the probability of choosing any of the $M$ most likely outcomes is weighted by the relative probabilities between these realizations. This sampling method is motivated by the finding that human decisions are likely to be based on the few most probable outcomes rather than on the full probability distributions \cite{li3computation}. The meta parameter $M$ can be tuned to a desired exploration/exploitation ratio. $M$=1 corresponds to sampling with very low temperatures (or argmax) and leads to the generation of periodic short motifs, probably due to the dynamics of the model entering a limit cycle. Then, the bigger the $M$ the more explorative are the generated MIDI sequences. After an empirical parameter tuning, we set $M$ to 3 as the authors believe this to be a good tradeoff between exploitation and exploration. 

BachProp has been implemented in Python using the Keras API \cite{chollet2015keras}. Code is available on GitHub\footnote{for the authors to remain anonymous, if accepted, the gitHub link will be revealed for the camera ready only \url{https://github.com/}}. 
\section{Results and discussion}
\label{sec:results}

In this section, we discuss the results of BachProp in terms of predictions, generation, and evaluation.

\subsection{Datasets}

\begin{table}
\caption{Description of data sets}
\label{tab:datasets}
\vskip 0.15in
\begin{center}
\begin{small}
\begin{sc}
\begin{tabular}{lcc}
\toprule
Dataset &  Size [score] & Size [note]\\
\midrule
Nottingham & 1035 & 313'890\\
Chorales &  381 & 56'383\\
String quartets & 215 & 738'739 \\
John Sankey & 135 & 358'211\\
Bach & 647 & 841'921\\
\bottomrule
\end{tabular}
\end{sc}
\end{small}
\end{center}
\vskip -0.1in
\end{table}

\begin{figure}[t]
\begin{tabular}{l}
\bf{A}\\
\includegraphics[width=\columnwidth]{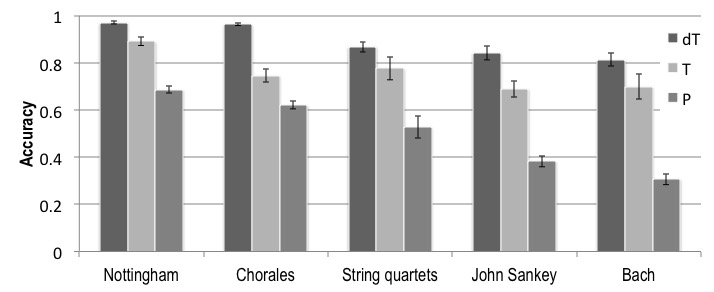}\\
\bf{B}\\
\includegraphics[width=\columnwidth]{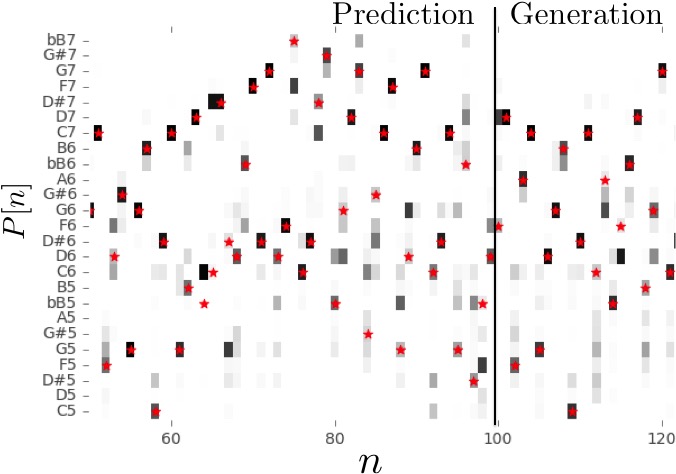}
\end{tabular}
\caption{\textbf{Accuracy, prediction and generation.} \textbf{A} The accuracy of predictions made by BachProp on unseen data for the three features $dT$, $T$, and $P$ for the five data sets. The accuracy is the fraction of correct predictions on an entire song. Mean and standard deviation of the accuracy are computed on the validation set. \textbf{B} Prediction and generation of a pitch sequence with BachProp. In greyscale (white =0 and black=1) is shown the network output $\mathbf{y}^P[n]\approx Pr(P[n+1])$. Until the 100th time step, BachProp is predicting the upcoming pitch (represented by red stars) from a Bach chorales. After the 100th time step, the red stars correspond to pitches that BachProp sampled from the network output distributions. Only a selection of active pitches is shown.}
\label{fig:results}
\end{figure}
We applied BachProp to five MIDI sequences corpora with different musical structures and styles (see Table \ref{tab:datasets}). The Nottingham database\footnote{\label{notchor}Nottingham and Chorales: \url{http://www-etud.iro.umontreal.ca/~boulanni/icml2012}} contains British and American folk tunes. The musical structure of all songs is very similar with a melody on top of simple chords. The Chorales corpus\footnoteref{notchor} includes hundreds of four-part chorales harmonized by Bach. Every chorale shares some common structures, such as the number of voices and rhythmical patterns. Therefore, we consider both Nottingham and Chorales corpora as homogeneous data sets. The John Sankey data set\footnote{\label{bach}John Sankey and Bach: \url{http://www.jsbach.net}} is a collection of MIDI sequences recorded by John Sankey on a digital keyboard. The sequences include works from Bach only but across the corpus pieces are rather different. In addition, this data set was recorded live from the digital keyboard and thus the temporal information needs normalization for efficient learning. The string quartets data set\footnote{String quartets: \url{hhttp://www.stringquartets.org}} includes string quartets from Haydn and Mozart. Here again, there is a large heterogeneity of pieces across the corpus. At last,
the Bach corpus\footnoteref{bach} contains all Bach pieces present on the Dave's J.S. Bach page. This includes many different pieces from solo organ works to the Cello Suites.

\subsection{Training}

The artificial neural network was trained using the procedure described in Section \ref{sec:training} on each corpus independently. The model architecture, the number of neurons and other hyper parameters are identical for all data sets. Because BachProp is not tuned on any particular corpus, it may be considered an universal algorithm of music composition. 

We observe the saturation of accuracies on both training and validation sets after less than 2000 epochs on all datasets. The accuracies after 2000 epochs are presented in Figure \ref{fig:results}\textbf{A}. We observe that the predictive performances of BachProp scales with the structure complexity. For homogeneous corpora with many examples of similar structures, BachProp can predict notes with accuracies as high as 90\% for timing, 80\% for duration and 60\% for pitch. For more heterogeneous data sets these performances decrease to 80\%, 70\%, and 30\%, respectively. To illustrate how prediction and generation are performed, Figure \ref{fig:results}\textbf{B} shows the pitch output of BachProp's note transition model.

However, these statistics do not inform about the quality of the music scores that BachProp generates. In the following sections, we discuss the end results of BachProp: new music scores generated by to the trained networks.

\subsection{Prediction and generation}

\begin{figure}[t]
\includegraphics[width=\columnwidth]{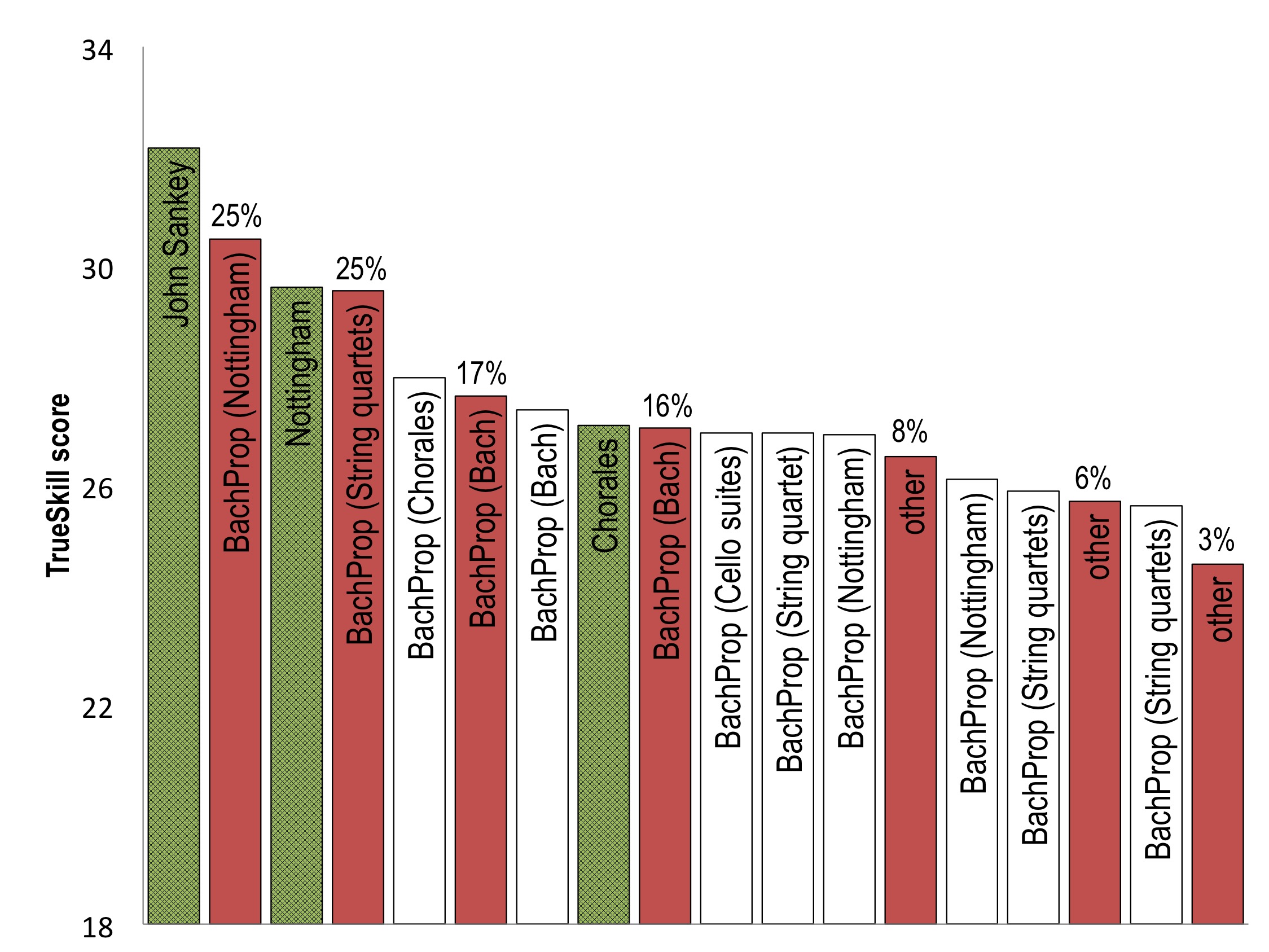}
\caption{\textbf{Results of the CrowdAI challenge on automated music composition.} Only the 18 top-rated MIDI submission are shown. The score is computed on the basis of 1756 pairwise comparisons from 90 MIDI submissions using the TrueSkill rating method. Out of the 90 submissions, 15 are new music scores generated by BachProp and 3 are a random selection of MIDI files from original corpora (green boxes with texture). The corpus on which BachProp was trained is written in brackets. \emph{other} stands for submissions coming from other (and unknown) algorithms than BachProp. Seven extracts from the submissions highlighted in red were chosen to be performed live by a string quartet. The audience was asked to select their preferred extract. The percentages above the boxes are the result of 356 individual votes. Each subject had a single vote to indicate the piece he/she liked best.}
\label{fig:evaluation}
\end{figure}


Renderings of note sequences generated by BachProp are available for listening on the webpage containing media for this paper\footnote{\label{media}Media webpage: \url{https://goo.gl/Xyx7WV}}. They are the results of BachProp after being trained on one of the five corpora. The length of each song can be infinite without losing any coherence as illustrated by online examples on the media webpage. We encourage readers to listen to these examples to convince themselves of the ability of BachProp to generate unique and heterogeneous new music scores. For better comparison, all examples are rendered using the same digital instrument. In addition, we also included examples from the original data sets. 

Professional musicians reported that music scores imagined by BachProp are very pleasant to listen. Even when the predictive performance of BachProp are lower as for the Bach data set, BachProp is able to produce convincing and coherent music pieces in any key. 

In the next section, we present a challenge designed to evaluate the subjective quality of music scores generated by BachProp.  

\subsection{Evaluation}
\label{evaluation}

Music scores generated by BachProp have been evaluated on the CrowdAI platform\footnote{\url{https://www.crowdai.org/}} as part of the AI-generated music challenge. The challenge was designed to rate MIDI files generated by different algorithms. Human evaluators were asked to give a preference between two 30 seconds extracts from 1 hour long MIDI sequences generated by music models. The challenge included control submissions (random selections of pieces from the John Sankey, Chorales and Nottingham corpora) as well as MIDI generated by other participants.

The scores of submissions were computed using the TrueSkill rating system  \cite{herbrich2007trueskill}. TrueSkill computes a score for each submission ensuring that top-rated submissions are highly preferred with high certainty. The results of the evaluation are presented in Figure \ref{fig:evaluation}. Controls apart, the 9 top-rated submissions were generated by BachProp. Surprisingly, the TrueSkill scores of original music corpora are undistinguishable from music generated by BachProp on the same corpora. 

Finally, a selection of seven extracts from top-rated submissions were arranged for a string quartet and performed live. After the performance, the audience was asked to select their preferred music piece. The video of the performance is available for readers on the media webpage\footnoteref{media}. The votes shown in Figure \ref{fig:evaluation} revealed a strong preference for two music scores generated by BachProp trained on respectively the Nottingham and String quartets data set. As an additional validation of the evaluation protocol, we observe that the votes from the live performance and ratings from the challenge are consistent across the two competitions. 

\section{Conclusion}

In this paper, we presented BachProp, an algorithm for general automated music composition. This generative model of music scores is able to compose in any style provided with examples of this style. We conducted an evaluation survey, which revealed that music generated by BachProp is not less preferred than music coming from the original corpus on which BachProp was trained. A central method of BachProp is its capacity to represent MIDI sequences in a low dimensional space. We developed this representation to be automatically applied to all MIDI sequences, even ones that were recorded from live performances. It allows BachProp to be trained on many new data sets and potentially on every possible MIDI sequences. Finally, BachProp was shown to be able to learn many different musical structures from heterogeneous corpora and create new music scores according to these extracted structures. For further directions, the authors propose to explore how repetitions, a feature inherent in music, can be better modeled by such algorithms. We believe that this feature is key for further improving neural models of music composition.

\section*{Acknowledgements}

We thank Johanni Brea and Samuel Muscinelli for helpful discussions, and Marcel Salath\'e and Sharada Prasanna Mohanty for the crowdAI challenge. Financial support was provided by \'Ecole Polytechnique F\'ed\'erale de Lausanne. 



\bibliography{BachProp}
\bibliographystyle{icml2018}

\end{document}